\documentclass{IEEEtran4PSCC}

\usepackage[noadjust]{cite}

\ifCLASSINFOpdf
   \usepackage[pdftex]{graphicx}
\else
   \usepackage[dvips]{graphicx}
\fi
\usepackage[cmex10]{amsmath}
\usepackage{algorithm}
\usepackage{algpseudocode}
\makeatletter
\let\old@ps@headings\ps@headings
\let\old@ps@IEEEtitlepagestyle\ps@IEEEtitlepagestyle
\def\psccfooter#1{%
    \def\ps@headings{%
        \old@ps@headings%
        \def\@oddfoot{\strut\hfill#1\hfill\strut}%
        \def\@evenfoot{\strut\hfill#1\hfill\strut}%
    }%
    \def\ps@IEEEtitlepagestyle{%
        \old@ps@IEEEtitlepagestyle%
        \def\@oddfoot{\strut\hfill#1\hfill\strut}%
        \def\@evenfoot{\strut\hfill#1\hfill\strut}%
    }%
    \ps@headings%
}
\makeatother

\psccfooter{%
        \parbox{\textwidth}{\hrulefill \\ \small{23rd Power Systems Computation Conference} \hfill \begin{minipage}{0.2\textwidth}\centering \vspace*{4pt} \includegraphics[scale=0.06]{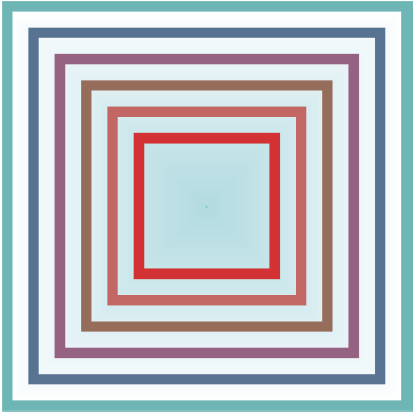}\\\small{PSCC 2024} \end{minipage} \hfill \small{Paris, France --- June 4 -- 7, 2024}}%
}

\newcommand{\dev}[1]{{{$\pm$ \scriptsize #1}}} 
\input{Definitions}

\usepackage{tikz}
\usepackage{multirow}
\usepackage{multicol}
\usepackage{booktabs}
\usepackage{subcaption}
\usepackage{amsmath}
\usepackage{mathtools}
\usepackage[hyphens]{url}
\usepackage{diagbox} 

\begin{document}
%

\title{Physics-Informed Heterogeneous Graph Neural Networks for DC Blocker Placement}


\author{\IEEEauthorblockN{Hongwei Jin\IEEEauthorrefmark{1},
    Prasanna Balaprakash\IEEEauthorrefmark{2},
    Allen Zou\IEEEauthorrefmark{3},
    Pieter Ghysels\IEEEauthorrefmark{3},
    Aditi S. Krishnapriyan\IEEEauthorrefmark{4, 3}, \\
    Adam Mate\IEEEauthorrefmark{5},
    Arthur Barnes\IEEEauthorrefmark{5}, and
    Russell Bent\IEEEauthorrefmark{6}}

  \vspace{3mm}
  \IEEEauthorblockA{\IEEEauthorrefmark{1}Mathematics and Computer Science Division \\
    Argonne National Laboratory, Lemont, IL}
  \vspace{1mm}
  \IEEEauthorblockA{\IEEEauthorrefmark{2}Computing and Computational Sciences Directorate \\
    Oak Ridge National Laboratory, Oak Ridge, TN}
  \vspace{1mm}
  \IEEEauthorblockA{\IEEEauthorrefmark{3}Computational Research Division \\
    Lawrence Berkeley National Laboratory, Berkeley, CA}
  \vspace{1mm}
  \IEEEauthorblockA{\IEEEauthorrefmark{4}
    University of California, Berkeley, CA}
  \vspace{1mm}
  \IEEEauthorblockA{\IEEEauthorrefmark{5}Information Systems and Modeling Group and \IEEEauthorrefmark{6}Applied Mathematics and Plasma Physics Group \\
    Los Alamos National Laboratory, Los Alamos, NM}
}


\maketitle

\begin{abstract}

  The threat of geomagnetic disturbances (GMDs) to the reliable operation of the bulk energy system has spurred the development of effective strategies for mitigating their impacts.
  One such approach involves placing transformer neutral blocking devices, which interrupt the path of geomagnetically induced currents (GICs) to limit their impact. The high cost of these devices and the sparsity of transformers that experience high GICs during GMD events, however, calls for a sparse placement strategy that involves high computational cost.
  To address this challenge, we developed a physics-informed heterogeneous graph neural network (PIHGNN) for solving the graph-based dc-blocker placement problem. Our approach combines a heterogeneous graph neural network (HGNN) with a physics-informed neural network (PINN) to capture the diverse types of nodes and edges in ac/dc networks and incorporates the physical laws of the power grid.
  We train the PIHGNN model using a surrogate power flow model and validate it using case studies. Results demonstrate that PIHGNN can effectively and efficiently support the deployment of GIC dc-current blockers, ensuring the continued supply of electricity to meet societal demands.
  Our approach has the potential to contribute to the development of more reliable and resilient power grids capable of withstanding the growing threat that GMDs pose.
\end{abstract}




\section{Introduction} \label{Sec:Introduction}

Geomagnetic disturbances (GMDs) pose a serious threat to the continuous and reliable operation of the bulk energy system \cite{NERC-HILF-2010}.
During severe space weather, charged and magnetized particles are blown away from the Sun. These disrupt the Earth's magnetic field and drive geomagnetically induced currents (GICs) -- quasi-dc currents produced by the induced low-frequency electric fields -- in the conductive infrastructure, which flow into the high-voltage network through the neutrals of transformers \cite{pirjola2000-gic, NERC2012-gmd}.
Depending on intensity, GICs can adversely impact transmission networks and equipment through overheating and thermal degradation in transformers, misoperation in protective devices, voltage collapses and the loss of reactive power support, and in the worst case, widespread infrastructure damage and tripping of transmission lines, leading to cascading power outages \cite{NERC2012-gmd, barnes21-hiddenfailures, mate21-pmsgmd-cascade, mate21-pmsgmd}.

A handful of strategies, for both system operation (transmission line switching, generation re-dispatch, and load-shedding) and system planning (placement of dc-blockers or series capacitors, and equipment stockpiling), have been explored over the years to mitigate the impacts of GMDs.
This paper focuses on the placement of transformer neutral blocking devices -- the installation of GIC dc-current blockers -- which consists of injecting a shunt capacitor in series with the grounding point of transformer neutrals to interrupt the path of GICs \cite{yang2019optimal}. Unlike other strategies, this approach does not reduce the power-carrying capacity of the grid, but may interfere with ground-fault relaying, which in turn requires additional switching devices and increased capital costs.
Given the high initial costs of these devices, and that the number of transformers that experience high GICs during GMD events is sparse \cite{horton_magnetohydrodynamic_2017}, sparse dc-blocker placement is necessary. This motivates the need for optimization-based placement formulations to guide the choices of where to place blocking devices.


Determining the optimal placement of GIC blocking devices is a challenging optimization problem formulated as a difficult mixed-integer non-linear program (MINLP). Developing effective techniques to solve such complex MINLP models for large-scale power systems has been an active research area for decades~\cite{molzahn2017survey, roald2023power, panciatici2014advanced}.

Recently, machine learning (ML) approaches have shown great promise for tackling these challenging optimization problems by training ML models to predict high-quality solutions efficiently~\cite{kody2022modeling, fioretto2020predicting, chen2022learning}, avoiding the need to solve the optimization directly. These ML surrogate models have demonstrated impressive performance on traditional power systems optimization tasks.

Motivated by these promising ML developments, this paper explores applying machine learning techniques to the critical problem of optimally placing GIC blocking devices to mitigate geomagnetically induced current impacts during solar storms. We develop a novel ML surrogate model approach that leverages recent advances in graph neural networks and physics-informed neural networks to effectively capture the complex network structure and physical laws governing power grid systems.

More specifically, the application of graph neural networks (GNNs) in power grid systems has been a rapidly developing area of research in recent years.
For example, \cite{owerko2020optimal} applied GNNs to learn a local and scalable parameterization of the optimal power flow (OPF) solution, while \cite{donon2020neural} learned to perform ac OPF computations by directly minimizing the violation of physical laws during training.
Additionally, \cite{liao2022short} proposed a new approach to predicting short-term power generation from renewable energy sources, combining a graph convolutional network and a long short-term memory model to capture both spatial and temporal correlations.
Furthermore, \cite{park2023confidence} presents a novel approach for solving reliability assessment commitments (RAC) optimization problems in power systems using GNNs and uncertainty quantification.

However, these models are designed for homogeneous graphs, which cannot fully capture the complex relationships and dependencies between different types of nodes and edges in ac/dc networks. The physics of GMD impacts on power system cause induced low-frequency voltages on electric transmission lines. A reasonable approximation is to model these as dc voltages. GICs can be calculated on transmission lines by generating a dc network from an ac transmission network model. In this dc network, the nodes correspond to both buses in the ac network and substation grounding grids, while edges correspond to transmission lines and transformer windings. This dc network is linked to the ac transmission network by reactive power loss resulting from half-cycle saturation on transformers. This is calculated as a function of the weighted sum of GICs flowing though each winding of a transformer and mapped to one of the transformer terminals \cite{overbye_integration_2012}. We will further discuss the details in Section~\ref{sec:h_graph}.

In addition, physics-informed neural networks \cite{raissi2019physics} have drawn attention in recent years to address the problem of learning from data while respecting the underlying domain. With such an approach, constraints can be added to the ML model to respect the underlying physics~\cite{raissi2019physics, cai2021physics, du2023neural}, or by combining more explicit information from numerical solvers with the ML model~\cite{lu2018beyond, chalapathi2024scaling}. In this work, we incorporate the physical information of the power grid into the ML training process.
This gap in the power grid literature motivates our research to explore the potential of heterogeneous graph neural networks in solving the dc-current blocker placement problem while ensuring the satisfaction of physical laws.


This paper proposes a novel heterogeneous graph neural network (HGNN) approach to model the complex ac/dc power grid network and predict optimal locations for placing GIC dc-current blockers to mitigate geomagnetic disturbance impacts. The HGNN captures the diverse node and edge types present in integrated ac/dc networks through tailored node and edge embeddings. Building upon this base HGNN model, we develop an innovative physics-informed heterogeneous graph neural network (PIHGNN) that incorporates the fundamental physical laws governing power grid operation. The PIHGNN combines the HGNN's predictions with a surrogate power flow model solver to ensure predictions respect the nonlinear ac/dc power flow constraints.

To the best of our knowledge, this is the first work that combines the powerful representational capabilities of graph neural networks for capturing network structure with physics-informed neural network techniques for enforcing physical law constraints. Applying this novel PIHGNN approach to the critical problem of optimizing GIC dc-current blocker placement is an important contribution. By addressing this gap between data-driven graph representations and physical power system models, our research enables more reliable and resilient power grid operation in the face of growing geomagnetic disturbance threats. The proposed PIHGNN model provides a cost-effective and computationally efficient machine learning surrogate for supporting real-world GIC blocker deployment decisions.

The rest of the paper is structured as follows:
In Section II., we introduce the mathematical formulation of blocker placement with objectives and constraints.
In Section III., we describe our ML surrogate approach and the methodology used to train and validate the model.
In Section IV., we present two case studies that demonstrate the effectiveness of our approach and discuss the results.
Finally, in Section V., we conclude with a brief discussion of the implications of our research and future directions for improving power grid resilience against GMDs with the use of neural networks.

\section{Problem Formulation} \label{sec:problem-formulation}

In this section, we discuss the GIC dc-current blocker placement formulation from an optimization problem perspective.
This formulation consists of four distinct pieces: the formulation used to model ac power flows in a transmission network, the formulation used to model GIC flows in a transmission network, the formulation that links GIC and ac formulations, and the formulation for modeling blockers.
It is important to note that while the ac and GIC flows are formulated for the same physical system, the formulations utilize unique representations of the system.

\subsection{AC Power Flow}

The ac power flow constraints consist primarily of the set of equations that represent the physics of power flow on an ac transmission network, and the set of inequalities that represent the operational limits of the network.
This is formulated as a MINLP maximum-load-delivered (MLD) formulation, as described in \cite{coffrin2018relaxations}.

\subsubsection{Objective}
\indent

\noindent
The problem objective is to maximize the amount of load delivered.
This can be alternately formulated as minimizing the cost associated with load-shedding:
\begin{align}
  \label{eq:load-shed-obj}
  \allowdisplaybreaks
  \mathrm{min.} \ \smashoperator{\sum_{i \in \mathcal{D}}} \abs{p_i^d} \kappa_i^d z_i^d
\end{align}
where
$\mathcal{D}$ is the set of loads in the ac network;
$p_i^d$ is the dispatched power of load $i$;
$\kappa_i^d$ is the cost of shedding load $i$; and
$z_i^d \in \{0,1\}$ is the load-shedding variable corresponding to load $i$.

\vspace{0.1in}
\subsubsection{Nodal power balance constraint}

\begin{align*}
  \label{eq:mls-kcl}
  \allowdisplaybreaks
  \small
  \sum_{j \in {\mathcal{E}_i^{a+}}}
   & \left( \mathbf{s}_j^+ - \mathbf{i} q_j^{loss+}\right) - \sum_{j \in {\mathcal{E}_i^{a-}}} \left(\mathbf{s}_j^-   + \mathbf{i} q_j^{loss-}\right) \\
   & = \sum_{j \in \mathcal{G}_i} \mathbf{s}_j^g - \sum_{j \in \mathcal{D}_i} z_j^d \mathbf{s}_j^d - \mathbf{y_i^s}\abs{\mathbf{v}_i}^2,
  \quad \forall i\in {\mathcal{N}^a}
\end{align*}
where
$\mathcal{N}^a$ is the set of ac buses;
$\mathcal{E}_i^{a+}$ is the set of branches flowing into bus $i$;
$\mathcal{E}_i^{a-}$ is the set of branches flowing out of bus $i$;
$\mathbf{s}_j$ is the complex power flow through branch $j$;
$q^{loss}_j$ is the reactive power loss of the branch resulting from GIC;
$\mathcal{G}_i$ is the set of generators connected to bus $i$;
$\mathbf{s}_k^g$ is the complex power produced by generator $k$;
$\Dcal_i$ is the set of loads connected to bus $i$;
$\mathbf{s}_k^d$ is the scheduled power of load $k$;
$z_k^d \in [0,1]$ is the load shed variable;
$\mathbf{y}_i^s$ is the shunt admittance of bus $i$; and
$\mathbf{v}_i$ is the voltage phasor at bus $i$.

\vspace{0.1in}
\subsubsection{Branch flow constraint}
\indent

\noindent $ \forall \; i \in {\mathcal{E}^a} $
\begin{subequations}
  \label{eq:ac-ohms-law-cons}
  \allowdisplaybreaks
  \small
  \begin{align}
    \mathbf{s}_i^+ & = \left( \mathbf{y}_i^* - \mathbf{i}\frac{\mathbf{b}_i^{sh}}{2} \right) \abs{\mathbf{v}_j}^2 - \mathbf{y}_i^*\mathbf{v}_j \mathbf{v}_k^* \\
    \mathbf{s}_i^- & = \left( \mathbf{y}_i^* - \mathbf{i}\frac{\mathbf{b}_i^{sh}}{2} \right) \abs{\mathbf{v}_k}^2 - \mathbf{y}_i^*\mathbf{v}_j^* \mathbf{v}_k
  \end{align}
\end{subequations}
\noindent where
${\mathcal{E}^a}$ is the set of ac edges;
$\mathbf{y}_i$ is the series admittance of branch $i$; and
$\mathbf{b}_i^{sh}$ is the line charging capacitance (equal to 0 for transformers).

\vspace{0.1in}
\subsubsection{AC Operational Limit Constraints}
\indent

\noindent
The voltage limits for each node (bus) $i \in {\mathcal{N}^a} $ are defined by,

\begin{equation}
  \allowdisplaybreaks
  \small
  \underline{v}_i \leq \abs{\mathbf{v}_i} \leq \overline{v}_i
  \label{eq:ac-voltage-bounds}
\end{equation}
\noindent where
the lower and upper bounds of the voltage magnitude at node $i \in \mathcal{N}^a$ are set by the range $[\underline{v}_i,\,\overline{v}_i]$.

\noindent
The thermal limits of an edge $i \in {\mathcal{E}^a}$ are defined by,

\begin{subequations}
  \allowdisplaybreaks
  \small
  \begin{align}
    \abs{\mathbf{s}_i^-} \leq \overline{s}_i \\
    \abs{\mathbf{s}_i^+}\leq \overline{s}_i
  \end{align}
  \label{eq:ac-branch-limit}
\end{subequations}
\noindent where
$\overline{s}_i$ is the thermal limit of edge $i$.

\noindent
Voltage stability limits on the phase angle difference on edge $i$, between nodes $j$ and $k$, are defined by,

\begin{equation}
  \allowdisplaybreaks
  \small
  \underline{\theta}_i \leq \arg(\mathbf{v}_k) - \arg(\mathbf{v}_j) \leq \overline{\theta}_i
  \label{eq:ac-voltage-angle-limit}
\end{equation}
\noindent where
$\underline{\theta}_i$ and $\overline{\theta}_i$ represent the upper and lower bounds on the phase angle difference.

\noindent
Given the set of generators $\mathcal{G}$, the capacity of power generation of $ i \in \mathcal{G} $ is defined by,

\begin{subequations}
  \label{eq:ac-gen-limit}
  \allowdisplaybreaks
  \small
  \begin{align}
    \underline{p}_i^g & \leq \mathcal{R}(\mathbf{s}_i^g) \leq \overline{p}_i^g \\
    \underline{q}_i^g & \leq \mathcal{I}(\mathbf{s}_i^g) \leq \overline{q}_i^g
  \end{align}
\end{subequations}
where $\mathcal{R}(\mathbf{s})$ and $\mathcal{I}(\mathbf{s})$ are the real and imaginary components of is the real part of $\mathbf{s}$, respectively, $\underline{p}_i^g$ and $\overline{p}_i^g$ represent the lower and upper bounds on real power for generator $i$; and
$\underline{q}_i^g$ and $\overline{q}_i^g$ represent the lower and upper bounds on real power for generator $i$.

\subsection{GIC Flow}

The dc power flow constraints consist of the set of equations that represent the physics of GIC flow on a dc representation of the transmission network.
For a node $i \in {\mathcal{N}^d}$, the nodal flow balance constraint is defined by,

\begin{equation}
  \allowdisplaybreaks
  \small
  \smashoperator{ \sum_{j \in \mathcal{E}^{d+}_i}}I_j - \smashoperator{\sum_{j \in \mathcal{E}^{d-}_i}} I_j = (1 - z_i^b)a_i^sV_i
  \label{eq:gic-kcl}
\end{equation}
\noindent where
$\mathcal{N}^d$ is the set of nodes in the dc network;
$\mathcal{E}_i^{d+}$ is the set of directed edges connected to node $i$ in the dc network that are oriented towards node $i$, while $\mathcal{E}_i^{d-}$ is the set of directed edges connected to node $i$ that are oriented against node $i$;
for the problem of maximizing load delivered, $z_i^b \in \{0,1\}$ for node $i$ is a parameter indicating if there is a blocker at $i$, which forces GIC flow into a transformer neutral to 0;
for the blocker placement problem, it is the binary placement variable for the blocker; and
finally, $V_i$ is the quasi-dc voltage at node $i$ and $a_i^s$ is the resistance to remote earth at $i$.

\vspace{0.1in}
Next, the dc Ohm's law constraint for $i \in \mathcal{E}^{d} $ is defined by,

\begin{equation}
  \allowdisplaybreaks
  \small
  I_i = a_i(V_j - V_k + V_i^{br})
  \label{eq:gic-ohms-law}
\end{equation}
\noindent where
$a_i$ is the conductance of dc edge $i$; and
$V_i^{br}$ is the voltage induced on dc edge $i$ by the GMD.

\subsection{GIC and AC coupling}

The notion of ``effective'' GIC is to calculate a weighted sum of currents through transformer windings that is proportional to the transformer core flux, and therefore to the GIC-induced reactive power losses of the transformer.
This varies for different transformer configurations. These are defined $ \forall i \in {\mathcal{E}^a} $ by,
\begin{equation}
  \label{eq:gic_effect_3_Idmag}
  \allowdisplaybreaks
  \small
  \widetilde I_i =
  \begin{cases}
    I_{i^H}                                                      & \text{if $i \in \mathcal{E}^\Delta$} \vspace{.1cm} \\
    {\frac{\alpha_i I_{i^H} + I_{i^L} }{\alpha_i}}               & \text{if $i \in \mathcal{E}^y$} \vspace{.1cm}      \\
    {\frac{\alpha_i I_{i^S} + I_{i^C} }{\alpha_i + 1}}           & \text{if $i \in \mathcal{E}^\infty$} \vspace{.1cm} \\
    I_{i^H} + \frac{I_{i^L}}{\alpha_i} + \frac{I_{i^T}}{\beta_i} & \text{if $i \in \mathcal{E}^3$} \vspace{.1cm}      \\
    0                                                            & \text{otherwise}
  \end{cases}
\end{equation}
\noindent where $\mathcal{E}^\Delta$ is the set of grounded-wye delta transformers, $\mathcal{E}^y$ is the set of grounded-wye-grounded-wye transformers, $\mathcal{E}^\infty$ is the set of autotransformers, $\mathcal{E}^3$ is the set of three-winding transformers,
$\alpha_i = V_{i^H}^b/V_{i^L}^b$ is the transformer turns ratio for the primary and secondary windings, while $\beta_i = V_{i^H}^b/V_{i^3}^b$ is the transformer turns ratio for the primary and tertiary windings for the case of three-winding transformers;
$\widetilde I_i$ is the ``effective'' GIC current on the dc edge;
$I_{i^H}$ is the high-side winding current and $I_{i^L}$ is the low-side for two- or three-winding transformers;
$I_{i^S}$ is the series winding current and $I_{i^C}$ is the common winding current for auto-transformers; and
$i^H=H_i^d,\, i^L=L_i^d,\, i^S=S_i^d, \, i^C=C_i^d,\, i^3=T_i^d$ are set index variables for transformer terminal nodes in the dc network.

\vspace{0.1in}
The magnitude of ``effective'' GIC
for $ \forall i \in \mathcal{E}^{\tau}$ is then defined by,

\begin{equation}
  \allowdisplaybreaks
  \small
  \overline{I}_i = \left|{\widetilde{I}_i}\right|
  \label{eq:gic-ieff}
\end{equation}

where $\mathcal{E}^{\tau} = \mathcal{E}^\Delta \cup \mathcal{E}^y \cup \mathcal{E}^\infty \cup \mathcal{E}^3$ is the set of all transformers.

\vspace{0.1in}
Last, the reactive power loss increase for a transformer resulting from GIC-induced half-cycle saturation
$ \forall i \in \mathcal{E}^{\tau} $ is defined by,
\begin{equation}
  \allowdisplaybreaks
  \small
  q^{loss+}_i =
  \begin{cases}
    \sqrt{\frac{2}{3}}\cdot\frac{S_i^b}{V_k^b} \cdot \abs{\mathbf{v}_j} K_i\overline{I}_{i}^d & \text{if $k$ = $i^H$} \\
    0                                                                                         & \text{otherwise}
  \end{cases}
  \label{eq:xfmr-reactpwrloss-pos}
\end{equation}

\begin{equation}
  \allowdisplaybreaks
  \small
  q^{loss-}_i =
  \begin{cases}
    \sqrt{\frac{2}{3}}\cdot\frac{S_i^b}{V_j^b} \cdot \abs{\mathbf{v}_j} K_i\overline{I}_{i}^d & \text{if $j$ = $i^L$} \\
    0                                                                                         & \text{otherwise}
  \end{cases}
  \label{eq:xfmr-reactpwrloss-neg}
\end{equation}

\noindent where
$j = H_i$ is the high-side node associated with transformer $i$;
$q_j^{loss}$ is the GIC-induced reactive power loss at the high-side node $j$;
$S_i^b$ is the base power of transformer $i$;
$V_j^b$ is the rated voltage at the high-side node $j$; and
$K_i$ is the GIC reactive power loss of transformer $i$.

\vspace{0.1in}
The MLD problem is formulated by minimizing Eq.~\eqref{eq:load-shed-obj} subject to constraints Eqs.~\eqref{eq:ac-ohms-law-cons} -- \eqref{eq:xfmr-reactpwrloss-neg} being satisfied.

\subsection{GIC DC-Current Blocker Placement}

The blocker placement problem takes on the same form as the MLD problem, however, the blocker placement parameters $z_i^b$ become binary decision variables and an additional constraint is added to limit the cost of blocker placement to within a budget:

\begin{equation}\allowdisplaybreaks
  \smashoperator{\sum_{i \in \mathcal{N}_b^d}} \kappa_i^b z_i^b \le \overline{c}_b
  \label{eq:dc-blocker-const-cons}
\end{equation}
\noindent where
$\kappa_i^b$ is the cost of placing a GIC dc-current blocker at node (bus) $i$; and
$\overline{c}_b$ is the total budget for blocker placement.

\vspace{0.1in}
The blocker placement problem is formulated by minimizing Eq.~\eqref{eq:load-shed-obj} subject to constraints Eqs.~\eqref{eq:ac-ohms-law-cons} -- \eqref{eq:dc-blocker-const-cons} being satisfied.


\section{Learning Model}
\indent

Given the mathematical formulation of blocker placement, in this section, we present our ML approach using HGNN to solve the GIC dc-current blocker placement problem as a binary classification problem with a physics-informed neural network model.

\subsection{Heterogeneous Graph}
\label{sec:h_graph}
\begin{figure}
  \centering
  \includegraphics[width=.6\linewidth]{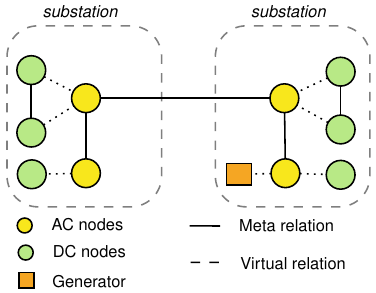}
  \caption{An abstract view of a heterogeneous graph for power grid (B4GIC).}
  \label{fig:h_graph}
\end{figure}

The power grid can be modeled as a heterogeneous graph, where different types of nodes represent different components of the grid (such as ac bus and generator) and the edges between nodes represent the connections between these components (such as transmission lines and transformers).
In addition to building a HGNN to predict blocker locations, we first consider the effects of GMD as quasi-dc current injection, and then we can use the ac-dc power flow model to solve the power flow problem.
Therefore, we also have a set of nodes and edges from the dc network, and a set of new edges between ac and dc networks.
We can use a combination of node and edge embeddings to represent the graph structure and node attributes (such as the type of power plant or the voltage level of a transmission line).
Figure~\ref{fig:h_graph} provides an abstract view of a heterogeneous graph for power grid B4GIC\footnote{PowerWorld simulator (B4GIC): https://www.powerworld.com/}, a typical small network studied for GIC problems.

\begin{definition}[Heterogeneous Graph]
  A heterogeneous graph is defined as an undirected graph $\Gds = (\Ncal, \Ecal, \Acal, \Rcal)$ where each node $v \in \Ncal$ and each edge $e \in \Ecal$ are associated with their type mapping function $\tau(v): \Vcal \rightarrow \Acal$ and $\phi(e): \Ecal \rightarrow \Rcal$, respectively.
\end{definition}

\paragraph*{\bf Meta Relation}
From an abstract view, a meta relation can be thought of as an edge $e=(s, t)$ that connects a source node $s$ to a target node $t$. The tuple $(s, e, t)$ represents the meta relation, where $e$ is the edge between the nodes, and $s$ and $t$ are the nodes themselves. The features associated with the meta relation are the attributes of the edge and the nodes it connects.
For example, the branch is an edge connecting two buses, denoted as (bus $s$, branch $e$, bus $t$), and has the attributes of the branch, the source bus, and the target bus.
By leveraging the features associated with meta relations, we can build more accurate and effective models that can be used to make predictions and decisions about the power grid.

\paragraph*{\bf Virtual Relation}
The power grid is hierarchical, and the meta relations alone are not sufficient to capture its hierarchical structure.
For instance, generators are a subset of buses, and they have attributes that are not shared with buses.
To address this limitation, we introduce virtual relations, a special type of meta relation that connects nodes without having edge features.
The virtual relation (bus $s$, conn $e$, gen $t$) represents a bus that is also a generator, i.e., bus $s$ and gen $t$ are the same physical entity, and the virtual edge conn $e$ is used to establish a connection between them.
Thus, this relation does not have any edge features but can still capture information from node to node.

\subsection{Heterogeneous Graph Neural Network}

Our approach uses HGNNs to learn the node and edge embeddings of a power grid.
Once the embeddings are learned, we use them to predict the likelihood of a blocker occurring in the dc buses, and identify the most effective locations to place blockers to mitigate the impact of GMDs.
Thus, we can use a HGNN to predict GIC dc-current blockers as a binary classification problem on dc buses.

To build a HGNN to predict blocker locations, we use a combination of node and edge embeddings to represent the graph structure and node attributes, such as a set of features of nodes (e.g., type, real demand, reactive demand, etc.) and a set of features of transmission lines (e.g., length, capacity, voltage level, etc.).

In general, the GNNs learn the embedding of each node by aggregating the information from its neighbors as follows,
\begin{align}
  \Hvec^t_v \leftarrow {\bf Aggr}_{\forall u \in N(v)} \rbr{f \rbr{\Hvec^{t-1}_u; \Hvec^{t-1}_v}}
\end{align}
where
$N(v)$ denotes the neighbors of $v$;
$\Hvec_v^t$ is the embedding of node $v$ at the $t$-th layer; and
$f$ is a learnable function given the node embeddings.
HGNNs can learn the node and edge embeddings simultaneously.

HGNN is a type of neural network designed to work with heterogeneous graph-structured data, and it has been shown to be effective in a variety of tasks, such as node classification, graph classification, and link prediction.
The key idea of HGNN is to iteratively update the node embeddings by aggregating information from neighboring nodes and learning a representation of the graph structure.
The update rule for the node embeddings can be written as:
\begin{align}
  \Hvec_i^{t+1} =
  \sigma\rbr{\sum_{j\in N(i)}\Hvec_j^t \Wvec_{ij}^{(\alpha)} + \sum_{j \in \Ncal(i)}\Evec_{ij}^t \Wvec_{ij}^{(\beta)}}
\end{align}
where
$\Hvec_i^t$ is the embedding of node $i$ at iteration $t$;
$\sigma$ is an activation function;
$N(i)$ is the set of neighboring nodes of $i$;
$\Wvec_{ij}^{(\alpha)}$ and $\Wvec_{ij}^{(\beta)}$ are weight matrices that capture the importance of the edge between nodes $i$ and $j$ for the node and edge embeddings, respectively; and
$\sum_{j\in N(i)}\Hvec_j^t \Wvec_{ij}^{(\alpha)}+\sum_{j\in N(i)}\Evec_{ij}^t \Wvec_{ij}^{(\beta)}$ is the aggregated feature representation of the neighbors of $i$.

In addition to the node embeddings, the model also learns edge embeddings to represent the transmission lines and transformers between nodes.
The edge embeddings are learned using a similar update rule:
\begin{align}
  \Evec_{ij}^{t+1}
  = \sigma\rbr{\sum_{k\in \Ncal(j)}\Evec_{kj}^t \Wvec_{ik}^{(\alpha)}+\sum_{k\in \Ncal(i)}\Hvec_k^t \Wvec_{ik}^{(\beta)}}
\end{align}
where
$\Evec_{ij}^t$ is the embedding of the edge between nodes $i$ and $j$ at iteration $t$.
For $t=0$, we take $\Hvec^0$ and $\Evec^0$ as the input node and edge features, respectively.

Once we have learned the node and edge embeddings, we use them to predict the likelihood of a blocker occurring in the power grid.
Specifically, we place blockers on the dc buses, where we define a probability distribution over the possible GIC blocker locations using a fully connected multi-layer perceptron (MLP) that takes the node and edge embeddings as input:
\begin{align}
  \PP(\zvec|\Hvec,\Evec) = \mathrm{MLP}(\Hvec,\Evec)
\end{align}
where
$\Hvec$ is the set of node embeddings of dc nodes;
$\Evec$ is the set of edge embeddings; and
$\zvec$ is the set of possible binary GIC blocker locations the model tries to predict.

The HGNN is trained using a variety of loss functions, such as cross-entropy loss between the predicted probability and the true labels, to optimize the prediction performance.
The learnable parameters of the HGNN include the weight matrices $\Wvec_{ij}^{(\alpha)}$ and $\Wvec_{ij}^{(\beta)}$ for the node and edge embeddings, and the parameters of the MLP for predicting the blocker locations, denoted as $\thetavec$ in general.
\subsection{Physics-Informed Heterogeneous GNN}

\begin{figure}
  \centering
  \includegraphics[width=\linewidth]{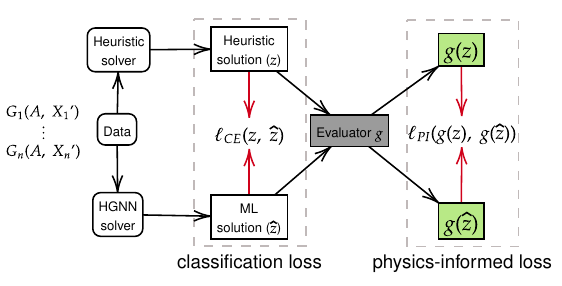}
  \caption{PIHGNN Framework}
  \label{fig:pihgnn_framework}
\end{figure}

Having location predictions simply provided by the HGNN is insufficient to solve the problem.
Since the core blocker placement problem is MINLP,
and a heuristic solver (ac-rect)~\cite{ryu2023heuristic} can only provide locally optimal solutions and not necessarily the global optimum, in this section, we introduce a novel approach that combines the strengths of HGNNs with the physics-based modeling of power grids.

The developed method, called physics-informed heterogeneous graph neural network (PIHGNN), leverages the predictive power of HGNNs with the physical laws governing power systems.
This allows for more accurate and reliable predictions of GIC dc-current blocker locations, which can help prevent equipment damage and power outages caused by GMDs.

PIHGNN is based on a re-evaluation step after reading out the blocker location predictions from the HGNN.
The predicted locations are sent into a surrogate problem, which measures the feasibility and optimality of the predictions in a real electric grid.
We denote the surrogate problem as function $\mathrm{g}(\cdot)$, which is a physics-based model that solves the MLD problem of a power grid given both ac and dc formulations; it takes the predicted blocker locations as input, and outputs a set of feasible and optimal values of the surrogate problem.

The surrogate model, $\mathrm{g}(\cdot)$, along with the HGNN, is trained using the backpropagation method.
The training data are the same as the HGNN problem with the additional output from the heuristic solver into the surrogate problem.
The total loss function was built not only from the HGNN problem but also from the surrogate problem as well, presented in Figure~\ref{fig:pihgnn_framework}, and is given
\begin{align}
  \ell(\Avec, \Xvec, \zvec) = \ell_{\mathrm{CE}} (\zvec, \hat{\zvec}) + \eta \ell_{\mathrm{PI}} \rbr{\mathrm{g}(\zvec), \mathrm{g}(\hat{\zvec})}
\end{align}
where $\Avec$ are the heterogeneous structural adjacencies, $\Xvec$ are the node and edge feature matrices, and $\eta$ is the regularization parameter;
$\ell_{\mathrm{CE}}(\cdot, \cdot)$ is the classification loss based on cross-entropy function from the HGNN problem; and
$\ell_{\mathrm{PI}}(\cdot, \cdot)$ is the physics-informed loss from the surrogate formulation.

\begin{figure}
  \centering
  \includegraphics{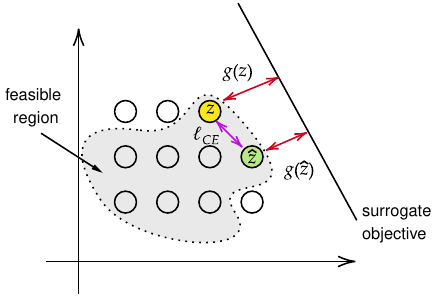}
  \caption{Evaluation of predictions. The heuristic solver provides the label $z$ (highlighted in yellow), and the GNN-based prediction $\hat{z}$ (highlighted in green) is evaluated by the evaluator $g$. The evaluator $g$ is a surrogate MLD model that takes the predictions from the heuristic solver and the GNN model as inputs.
  }
  \label{fig:mip}
\end{figure}

Figure~\ref{fig:mip} illustrates the rationale behind the loss function in PIHGNN.
A classification loss, typically based on a cross-entropy function, only ensures that the solution obtained from the HGNN ($\hat{\zvec}$) is close to the observed data ($\zvec$).
Due to the non-convex and non-smooth optimization of the blocker placement problem, the observed data is not the only solution to the original problem.
Moreover, the prediction from the HGNN model is not guaranteed to be a feasible solution to the original blocker placement problem.
To capture the underlying physical laws governing the behavior of power systems, we propose to use a surrogate model $\mathrm{g}(\cdot)$ to evaluate the quality of predictions, both in feasibility and optimality.
A physics-informed loss $\ell_{\mathrm{PI}}(\cdot, \cdot)$ offers a promising approach to simplify the complexity of the underlying physical laws but still serves as a surrogate measurement to the quality of solutions.

The advantages and contributions of PIHGNN are:
\begin{itemize}
  \item {\bf Improved accuracy}: PIHGNN combines the predictive power of HGNNs with the physical laws governing power systems, leading to more accurate predictions of blocker locations.
  \item {\bf Increased reliability}: The re-evaluation step using the surrogate model ensures that the predicted GIC dc-current blocker locations are feasible and optimal, reducing the risk of equipment damage and power outages caused by GMDs.
  \item {\bf Flexibility}: PIHGNN can be adapted to different scenarios and can handle various types of uncertainties, such as uncertainty in the blocker locations, uncertainty in the grid topology, and uncertainty in the grid parameters.
\end{itemize}

\begin{algorithm}
  \caption{Adaptive training of PIHGNN}\label{alg:adaptive_training}
  \begin{algorithmic}[1]
    \State \textbf{Data:} $\Gcal_1 (\Avec, \Xvec_1, \zvec_1), \cdots, \Gcal_n (\Avec, \Xvec_n, \zvec_n)$
    \State \textbf{HP:} $\eta$, $p$, $lr$, epoch, etc.
    \Procedure{Training}{$\Gcal_1, \cdots, \Gcal_n$}
    \For{epoch}
    \For{sample $\Gcal_i$ in batch}
    \State $\hat{\zvec}_i \gets$ HGNN($\Gcal_i$) \Comment{feedforward}
    \State $\ell_{\mathrm{CE}, i} \gets CE(\zvec_i, \hat{\zvec}_i)$ \Comment{classification loss}
    \EndFor
    \If{epoch \% p = 0}
    \For{sample $\Gcal_i$ in batch}
    \State $\ell_{\mathrm{PI},i} \gets d(g(\hat{\zvec}_i), g(\zvec_i)))$ \Comment{PI loss}
    \EndFor
    \EndIf
    \State $\ell = \smallfrac{1}{b} \sum_i (\ell_{CE, i} + \smallfrac{\eta}{\sqrt{epoch}} \ell_{PI, i})$
    \State Update $\thetavec$ \Comment{backpropagation}
    \EndFor
    \State Return HGNN with parameters $\thetavec$
    \EndProcedure
  \end{algorithmic}
\end{algorithm}

\vspace{-1em}
\section{Case Studies}

\subsection{Dataset}

Extensive experiments were conducted on two widely used synthetic test systems, EPRI21 and UIUC150, which are publicly available with PowerModelsGMD\footnote{https://github.com/lanl-ansi/PowerModelsGMD.jl} (PMsGMD) \cite{mate21-pmsgmd}.
The EPRI21 dataset is a 20-bus network with 7 generators and 31 branches, and the UIUC150 dataset is a 150-bus network with 27 generators and 218 branches.
Details of the PMsGMD data format are summarized in Table~\ref{tab:features}: tables ``\texttt{bus}'', ``\texttt{gen}'', and ``\texttt{gmd\_bus}'' contain the nodes in the heterogeneous graph; tables ``\texttt{branch}'', ``\texttt{branch\_gmd}'', and ``\texttt{gmd\_branch}'' contain the meta relations with a set of specific features; and tables ``\texttt{bus\_conn\_gen}'' and ``\texttt{gmd\_bus\_attach\_bus}'' contain the virtual relations without edge features.

\paragraph*{Data preprocessing}
To demonstrate the effectiveness of our proposed models, we first generate a set of synthetic data representing the GMD scenarios: 50 samples for each combination of E-field direction in $(45, 135)$ degree and magnitude $(5, 10, 15, 20)$, along with a uniformly random scalar perturbation in $[0.8, 1.2]$ for active and reactive demands on load buses.
With a total 400 samples for each dataset, we split the dataset into 80\%, 10\%, and 10\% for training, validation, and testing, respectively. We also normalize the node and edge features to $[0, 1]$ in order to stabilize the training process\footnote{The source code is accessible: \text{https://github.com/cshjin/swmp\_ml/}}.

\begin{table*}
  \centering
  \caption{Statistics of networks}
  \label{tab:features}
  \resizebox*{\textwidth}{!}{
    \begin{tabular}{c|c|cccccccc}
      \toprule
      (\# entries, \# features) & blocker locations & bus       & gen      & gmd\_bus & branch    & branch\_gmd & gmd\_branch & bus\_conn\_gen & gmd\_bus\_attach\_bus \\
      \midrule
      EPRI21                    & (8, )             & (19, 15)  & (7, 20)  & (27, 3)  & (31, 11)  & (31, 18)    & (37, 4)     & (7, 0)         & (27, 0)               \\
      UIUC150                   & (98, )            & (150, 15) & (27, 20) & (248, 3) & (218, 11) & (218, 18)   & (250, 4)    & (27, 0)        & (248, 0)              \\
      \bottomrule
    \end{tabular}
  }
\end{table*}

\subsection{Performance}

Formulated as a binary classification problem, we evaluated the performance of the proposed models on the test systems using the accuracy and ROC-AUC metrics with 10-fold cross validation,
where the accuracy is the ratio of the number of correct predictions to the total number of predictions, and the ROC-AUC is the area under the receiver operating characteristic curve, which is typically used to evaluate the performance of binary classification models.
We use 4 layers of heterogeneous transformer layer with 4 heads, and 4 layers of MLP with 128 hidden dimensions.
We apply the Adam optimizer \cite{KingmaB14} with a learning rate of 0.001 and weight decay of 0.0001 for the backpropagation with 200 epochs.
Table~\ref{tab:pred_perf} summarizes the prediction performance of the two models, HGNN and PIHGNN, on the two datasets.
Results indicate that PIHGNN outperforms HGNN on both datasets, with average accuracy improvements of 0.052 and 0.013, and average ROC-AUC improvements of 0.046 and 0.007, respectively. Regarding the performance variations of the datasets, EPRI21 has generally higher accuracy and ROC-AUC scores than UIUC150, which is primarily due to the complexity of UIUI150 being a larger system.
Results also suggest that PIHGNN is a more reliable and robust model than HGNN, especially when applied to a diverse and complex power system dataset.

\begin{table}[]
  \centering
  \caption{Prediction performance}
  \label{tab:pred_perf}
  \begin{tabular}{c|c|c|c}
    \toprule
    Dataset                  & Model  & Accuracy        & ROC-AUC         \\
    \midrule
    \multirow{2}{*}{EPRI21}  & HGNN   & 0.732 \dev{.08} & 0.755 \dev{.11} \\
                             & PIHGNN & 0.784 \dev{.12} & 0.801 \dev{.10} \\
    \midrule
    \multirow{2}{*}{UIUC150} & HGNN   & 0.702 \dev{.07} & 0.588 \dev{.07} \\
                             & PIHGNN & 0.715 \dev{.10} & 0.594 \dev{.12} \\
    \bottomrule
  \end{tabular}
\end{table}

\begin{table}[]
  \centering
  \caption{Efficiency comparison. The training time is the time per data sample to train the model, and the testing time is the time per data sample to evaluate the model on the test set. The heuristic solver is the time per data sample to solve the problem in Julia.}
  \label{tab:efficiency}
  \begin{tabular}{c|c|c|c}
    \toprule
    Dataset                  & Model            & Training (sec.) & Testing (sec.) \\
    \midrule
    \multirow{3}{*}{EPRI21}  & HGNN             & 4.75            & $\ll 1$        \\
                             & PIHGNN           & 7.50            & 2.31           \\
                             & Heuristic Solver & 54.056          & 54.12          \\
    \midrule
    \multirow{3}{*}{UIUC150} & HGNN             & 8.12            & $\ll 1$        \\
                             & PIHGNN           & 10.30           & 3.55           \\
                             & Heuristic Solver & 384.52          & 390.41         \\
    \bottomrule
  \end{tabular}
\end{table}

One advantage of the proposed HGNN/PIHGNN model is that it can be trained efficiently in a supervised manner.
To demonstrate this, Table~\ref{tab:efficiency} compares the efficiency of three methods -- HGNN, PIHGNN, and a heuristic solver~\cite{ryu2023heuristic} -- for solving the blocker placement problem.
Training and testing times for each method, expressed in seconds per data sample, is displayed; training time is the time required to train the model on the training set, while testing time is the time required to evaluate the model with one sample on the test set.
The heuristic solver~\cite{ryu2023heuristic},
is a baseline method that uses a stochastic learning optimizer which is implemented in Julia, and its time is also shown per data sample.
Table~\ref{tab:efficiency} shows that HGNN and PIHGNN have significantly lower training and testing times than the heuristic solver, with HGNN being the fastest method overall.
In particular, HGNN has training of 4.75 and 8.12 seconds per data sample on average from EPRI21 and UIUC150, respectively.
While on the testing set, HGNN has a negligible time of less than 1 second per data sample.
In contrast, the heuristic solver has much longer times, ranging from 1 minute to 10 minutes per data sample in EPRI21 and UIUC150 respectively.
These results indicate that both HGNN and PIHGNN are significantly more efficient than the heuristic solver, and are therefore more suitable for solving the blocker placement problem in large networks where the heuristic solver is time-consuming.

\begin{figure}
  \centering
  \includegraphics[width=.8\linewidth]{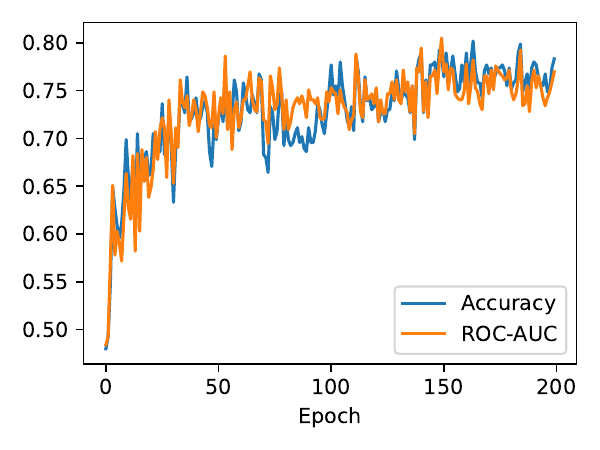}
  \caption{Training PIHGNN model on EPRI21}
  \label{fig:train_epri21}
\end{figure}

\begin{figure}
  \centering
  \includegraphics[width=.8\linewidth]{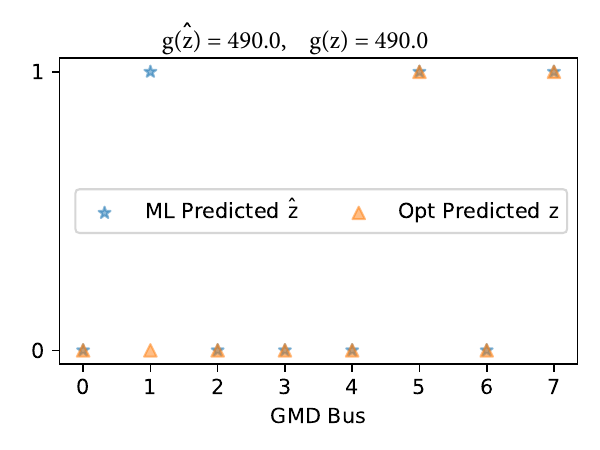}
  \caption{A surrogate evaluator for EPRI21}
  \label{fig:surrogate_eval}
\end{figure}

Figure~\ref{fig:train_epri21} shows the training process of the PIHGNN model on the EPRI21 dataset. Both accuracy and ROC-AUC scores are shown as a function of the number of training epochs.
The figure shows that the model achieves a high accuracy and ROC-AUC score over 0.75, respectively, after 150 epochs.
These results indicate that the PIHGNN model can be trained efficiently on the EPRI21 dataset, and can achieve a high prediction performance after a small number of training epochs.
More interestingly, Figure~\ref{fig:surrogate_eval} shows the surrogate evaluator on a single case in EPRI21. As shown, the PIHGNN prediction is slightly different from the heuristic solver, but the output from both solutions is exactly the same in the surrogate evaluator. This demonstrates the need for the surrogate evaluator to evaluate the PIHGNN prediction despite the binary labels from the HGNN model when compared with the MIP solver.

\begin{figure}
  \centering
  \includegraphics[width=.8\linewidth]{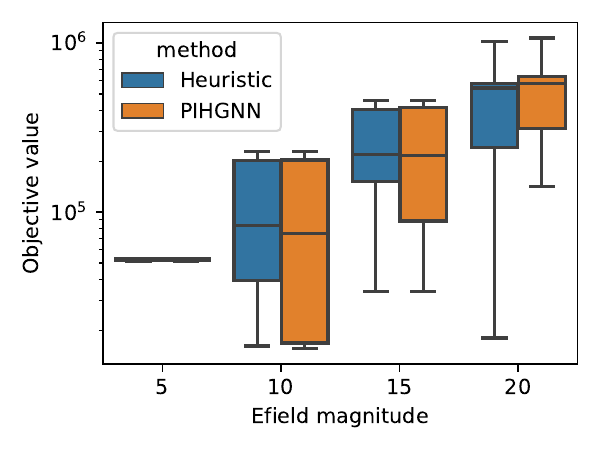}
  \caption{Objective function values w.r.t. E-field magnitude on EPRI21}
  \label{fig:physical-reeval}
\end{figure}

We further investigated the impact of the electric field (E-field) on the blocker placement problem by evaluating the PIHGNN model's predictions under different E-field magnitudes. Figure~\ref{fig:physical-reeval} depicts the objective function values for the PIHGNN model, applied to the EPRI21 dataset, as a function of increasing E-field magnitude (while fixing the direction at 45°). Notably, both the heuristic approach and the ML approach exhibit an increasing trend in objective function value with rising E-field magnitude. However, the ML approach achieves comparable objective function values to the heuristic approach, demonstrating similar average performance.

It is important to acknowledge that the non-convex and non-smooth nature of the optimization problem precludes exact equivalence between the objective function values of the PIHGNN model and the heuristic solver. Nevertheless, the observed closeness in their average values across various E-field magnitudes suggests the PIHGNN model's efficacy in predicting the objective function value of the blocker placement problem, which consequently allows for assessment of prediction quality. This aligns with the findings presented in Table~\ref{tab:efficiency}, emphasizing the suitability of the PIHGNN model for large-scale power systems, where the computational efficiency of training the model outweighs the substantial time required for the heuristic solver.
%

Last, we examined the generalization capability of our machine learning model by training it on one network and predicting on another.
This approach leverages the knowledge acquired during training on known datasets to make predictions on similar datasets, even with different network sizes, thanks to the graph neural networks (GNNs).
Such generalization ability is particularly useful in power systems where a model can learn patterns from one network and apply them to another without the need for explicit heuristics or simulated data.

Table~\ref{tab:generalization} shows the empirical results of training HGNN model (without true labels to be evaluated in the surrogate) on one network and testing it on another.
The model trained on EPRI21 data and tested on UIUC150 achieved an accuracy score of 0.672, which is lower than the model trained and tested on its own (0.718).
Similarly, the model trained on UIUC150 and tested on EPRI21 had an accuracy score of 0.661, which is lower than the model trained from its own data.
These results provide empirical evidence that a model can generalize to new data without requiring explicit labels or training data from the new network.

\begin{table}[]
  \centering
  \caption{Generalization of learned HGNN (accuracy scores reported)}
  \label{tab:generalization}
  \begin{tabular}{c|c|c}
    \toprule
    \diagbox[width=5em,height=2em]{\multirow{1}{*}{Train}}{\multirow{1}{*}{Test}}
            & EPRI21 & UIUC150 \\
    \midrule
    EPRI21  & 0.741  & 0.672   \\
    UIUC150 & 0.661  & 0.718   \\
    \bottomrule
  \end{tabular}
\end{table}

\section{Conclusion}

We introduced a physics-informed heterogeneous graph neural network (PIHGNN) for solving the graphically constrained blocker placement problem in power systems.
The proposed method combines the strengths of graph neural networks (GNNs) and physics-informed neural networks (PINNs) to efficiently and accurately optimize the placement of geomagnetically induced currents (GICs) dc-current blockers.
The GNN captures the complex relationships and dependencies between different types of nodes and edges in ac/dc networks, while the PINN ensures that the solution satisfies the physical laws of the power grid.
Further work will extend the proposed method to solve other optimization problems on power systems, such as voltage stability in presence of GICs, investigate other machine learning techniques, such as reinforcement learning, and incorporate spatiotemporal data for explicit physics-informed neural networks in the future.

\section*{Acknowledgements}

The work was jointly funded by the U.S. Department of Energy's Office of Electricity Advanced Grid Modeling program and the U.S. Department of Energy's Office of Science Scientific Discovery Through Advanced Computing (SciDAC) program, under the project ``Space Weather Mitigation Planning'' (contract number DE-AC02-06CH11357).
LA-UR-23-31427 -- approved for public release; distribution is unlimited.


\bibliographystyle{unsrt}
\bibliography{ref.bib}

\end{document}